\documentclass[aip,pop,reprint,a4paper]{revtex4-1}
\usepackage{cancel}
\usepackage{mathtools}
\usepackage{amsmath}
\usepackage{amssymb}
\usepackage{graphicx}% Include figure files
\usepackage{dcolumn}% Align table columns on decimal point
\usepackage{bm}% bold math

\newcommand{\p}{\text{p}}
\newcommand{\G}{\mathsf{G}}
\newcommand{\C}{\mathsf{C}}
\newcommand{\D}{\mathsf{D}}
\newcommand{\lp}{\langle\ell_\p\rangle}
\newcommand{\barx}{\bar{\bm{x}}}
\newcommand{\calG}{\mathcal{G}}
\newcommand{\calH}{\mathcal{H}}
\newcommand{\px}{\frac{\delta}{\delta\bm{x}}}
\newcommand{\pp}{\frac{\delta}{\delta\bm{p}}}
\newcommand{\func}[2]{\frac{\delta #1}{\delta #2}}
\newcommand{\bb}[1]{\mathbb{#1}}
\newcommand{\balpha}{\bm{\alpha}}
\newcommand{\bbeta}{\bm{\beta}}

\draft % marks overfull lines with a black rule on the right

\begin{document}

%\preprint{}

\title{Finite-dimensional collisionless kinetic theory} %Title of paper



\author{J. W. Burby}
 \affiliation{Courant Institute of Mathematical Sciences, New York, New York 10012, USA}
%\thanks{}


\date{\today}

\begin{abstract}
A collisionless kinetic plasma model may often be cast as an infinite-dimensional noncanonical Hamiltonian system. I show that, when this is the case, the model can be discretized in space and particles while preserving its Hamiltonian structure, thereby producing a finite-dimensional Hamiltonian system that approximates the original kinetic model. I apply the general theory to two example systems: the relativistic Vlasov-Maxwell system with spin, and a gyrokinetic Vlasov-Maxwell system. 

\end{abstract}

\pacs{}% insert suggested PACS numbers in braces on next line

\maketitle %\maketitle must follow title, authors, abstract and \pacs

\section{Introduction}
A basic approach to the study of continuum mechanical models is truncation. In the truncation process, continuous fields appearing in the model of interest are expressed in terms of some Galerkin basis, and then all but a finite number of basis functions are discarded. The resulting finite-dimensional dynamical system is often much simpler to study both analytically and numerically. A particularly famous outcome of applying truncation to a continuum fluid model was the discovery by Lorenz\cite{Lorenz_1963} of the strange attractor bearing his name. More generally, truncation is used often to develop computer simulations. When it is, the tendency is to use the term``semi-discretization" or ``continuous-time discretization" instead of truncation. A truncated model is semi-discrete because the fields are represented by a discrete (and finite) collection of degrees of freedom, while the time is represented by variable that assumes every value in $\mathbb{R}$.

Recently, a variety of authors\cite{Shadwick_2014,Evstatiev_2014,Xiao_2015,Qin_nuc_2016,He_2016,Kraus_2016_arXiv} have formulated truncations for collisionless kinetic plasma models that are compatible with Hamiltonian mechanics (see also Ref.\onlinecite{Squire_PIC_2012} for a compatible fully-discrete Vlasov-Maxwell system.) Here compatibility refers to the fact that the truncated models are finite-dimensional Hamiltonian systems with Hamiltonians and Poisson brackets inherited from the corresponding continuum theory. These truncations have already proven useful in the enterprise of developing structure-preserving kinetic simulations. In the future, there is great potential for squeezing even more utility from them. One tantalizing prospect is to use the Liouville measure associated with the finite-dimensional Hamiltonian structure for equation of state computations. Another is to use low-dimensional variants of these truncations as simplified models of nonlinear kinetic plasma behavior.

The purpose of this article is to present a significant generalization and unification of the previous Hamiltonian truncations for collisionless kinetic plasma models. By starting from a very general continuum kinetic model, the truncation developed here will simultaneously provide truncations for a variety of systems, including Vlasov-Maxwell, relativistic spin Vlasov-Maxwell, drift-kinetic Vlasov-Maxwell, and gyrokinetic Vlasov-Maxwell. The inclusion of the gyrokinetic Vlasov-Maxwell system here is especially significant because Hamiltonian truncations have not been found previously for this system. Aside from being manifestly Hamiltonian, the general truncation will also include two special features that have been identified as being important by previous authors. First, by using a field discretization as in Refs.\,\onlinecite{Xiao_2015,He_2016,Kraus_2016_arXiv}, gauge invariance and charge conservation will be preserved exactly. Second, by using the macroparticle formalism from Refs.\,\onlinecite{Shadwick_2014,Xiao_2015}, the truncation will assume the form of a low-noise particle-in-cell (PIC) discretization.

The article is organized as follows. In Section \ref{continuum}, I present the general continuum kinetic model and its Hamiltonian structure. In Section \ref{finite}, I construct a discretization of the general continuum model and present its Hamiltonian structure. As a first example,  in Section \ref{rel_spin}, I construct a new continuous-time discretization of the relativistic spin Vlasov-Maxwell system and present its Hamiltonian structure. Incidentally, this section's presentation of the Hamiltonian structure for the continuum relativistic spin Vlasov-Maxwell system is new. As a second example, I present a new continuous-time discretization of a gyrokinetic Vlasov-Maxwell system and present its Hamiltonian structure in Section \ref{gkvm}. 

%Why is it important that collisionless kinetic plasma models have Hamiltonian structure? One response might be ``conservation laws." Hamiltonian structure can be used to efficiently identify a Hamiltonian system's conservation laws. This response has some merit. Conservation laws in certain systems (e.g. gyrokinetics\cite{Scott_2010}) can be very hard to find by guessing. On the other hand, the response also has a serious weakness. In principle, conservation laws can always be found without knowledge of a system's Hamiltonian structure. A stronger, and more nuanced response would be ``reduced models." After discovering the Hamiltonian structure of a kinetic plasma model, approximation schemes can be applied directly to that structure. The resulting reduced model will itself be a Hamiltonian system, provided a modicum of care is taken. In particular, assuming symmetries connected with conservation laws are compatible with the approximation scheme, the reduced models will retain exact conservation laws. A great illustration of this point is found in Littlejohn's Hamiltonian formulation of guiding center theory.\cite{Littlejohn_1981} On the other hand, reduced kinetic models with exact conservation laws can also be found without using Hamiltonian structure, as is done for instance in the multiscale gyrokinetic theory of Ref.\,\onlinecite{Abel_2013}. Thus, this second response may still fail to convince some.

%%%%%%%%%%%%%%%%%%%%%%%%%%%%%%%%%%%%%%%%%%%%%%%%%%%%%%%%%%%%%%%%%%%
\section{The Hamiltonian structure of collisionless kinetic theory\label{continuum}}
In this section I will present a general collisionless electromagnetic kinetic plasma model with Hamiltonian structure. It will generalize all fully-electromagnetic kinetic models that have appeared in the literature. By constructing such a general model, I will be able to give a very general prescription for finding continuous-time finite-dimensional kinetic models with Hamiltonian structure. Continuous-time discretization of the continuum model presented in this section will be discussed in Section \ref{finite}.

%Before discussing the benefits of Hamiltonian structure in collisionless kinetic theory, it is sensible to display that structure explicitly. The purpose of this section is to do just that. This section will formulate a very general collisionless kinetic plasma model and present its Hamiltonian structure. The model will be general enough to reproduce the Vlasov-Maxwell system, the relativistic Vlasov-Maxwell system with spin, the drift-kinetic Vlasov-Maxwell system, and the gyrokinetic Vlasov-Maxwell system as special cases. The results contained in this section represent a significant generalization of those found in Refs.\,\onlinecite{Morrison_lifting_2013} and \onlinecite{Burby_gvm_2015}.

All kinetic plasma models prescribe the evolution law for a distribution function $F$ defined on a phase space $M$ of dimension $n$. On the space $M$, the dynamical equations governing the motion of an individual plasma particle assume the form of a first-order ordinary differential equation. The function $F$ counts plasma particles in subregions $U\subset M$ of $M$. Specifically, the number of particles in $U$ is given by integrating $F$ over $U$.

For the sake of concreteness, introduce coordinates $\bm{z}\equiv(\bm{x},\bm{m})$ on $M$. The component functions $\bm{x}\in\mathbb{R}^3$ and $\bm{m}\in\mathbb{R}^m$, where $m=n-3$, will be assumed to have the following interpretations. The value of $\bm{x}$ gives the spatial location of a particle (or quasi-particle). Space, which will be denoted $Q$, will be either infinite or periodic. The value of $\bm{m}$ gives a particle's ``internal state."  In Vlasov-Maxwell kinetic theory, $\bm{m}=\bm{v}$ is the particle velocity. In drift kinetic and gyrokinetic theory, $\bm{m}=v_\parallel$ is the guiding center or gyrocenter parallel velocity. More generally, $\bm{m}$ may encode the orientation of a composite particle, or even a particle's spin.

In the absence of collisions, the motion of a given particle through the phase space $M$ is governed by a single-particle Lagrangian $\ell$. A general form for this Lagrangian that encompasses all known Hamiltonian kinetic models is given as follows. Let $(\bm{A},\varphi)$ be the electromagnetic potentials, and $(\bm{E},\bm{B})$ the corresponding electromagnetic fields. Set $z^i=x^i$ when $i=1,2,3$ and $z^i=m^{i-3}$ when $i=4,...,n$. The Lagrangian $\ell$ is given by
\begin{align}\label{sin_part}
\ell=\theta_i(\bm{z};\bm{B})\dot{z}^i-K(\bm{z};\bm{E},\bm{B})+\frac{e}{c}\bm{A}(\bm{x})\cdot\dot{\bm{x}}-e\varphi(\bm{x}),
\end{align}
where $e$ is the particle's charge. The $n$ functions $\theta_i$ depend explicitly on the phase space coordinates $\bm{z}$ and parametrically on the magnetic field $\bm{B}=\nabla\times\bm{A}$. The dependence on $\bm{B}$ is in general nonlinear and nonlocal. Likewise, the function $K$ depends explicitly on $\bm{z}$ and parametrically on both the magnetic field and the electric field $\bm{E}=-c^{-1}\partial_t\bm{A}-\nabla\varphi$. Again, the field dependence is generally nonlinear and nonlocal. 

The total Lagrangian for all plasma particles is given by 
\begin{align}
L_{\chi}=&\sum \int (\theta_i(\bm{z};\bm{B})V^i(\bm{z})-K(\bm{z};\bm{E},\bm{B}))\,F(\bm{z})\,d^n\bm{z}\nonumber\\
+&\frac{1}{c}\int \bm{J}_f(\bm{x})\cdot\bm{A}(\bm{x})\,d^3\bm{x}-\int\rho_f(\bm{x})\varphi(\bm{x})\,d^3\bm{x},
\end{align}
where $\sum$ denotes a species sum. The functional $L_\chi$ can be thought of as the sum-over-particles of the single-particle Lagrangian $\ell$. The vector field $\bm{V}=V^i\partial_i$ on $M$ is the Eulerian phase space flow velocity. The $\bm{x}$-component of $\bm{V}$ gives a particle's $\bm{x}$-velocity $\bm{u}$, while the $\bm{m}$-component of $\bm{V}$ gives a particle's $\bm{m}$-velocity $\bm{w}$. Specifically, $V^i=u^i$ when $i=1,...,3$ and $V^i=w^{i-3}$ when $i=4,...,n$. The ``free" current and charge densities are given in terms of $F$ and $\bm{u}$ as
\begin{align}
\bm{J}_f(\bm{x})&=\sum e \int \bm{u}(\bm{x},\bm{m})\,F(\bm{x},\bm{m})\,d^m\bm{m}\\
\rho_f(\bm{x})&=\sum e \int F(\bm{x},\bm{m})\,d^m\bm{m}.
\end{align}

The general Lagrangian $L$ for the entire field-particle system that comprises an electromagnetic kinetic theory is given by
\begin{align}
L=L_{\chi}+\frac{1}{8\pi}\int (|\bm{E}(\bm{x})|^2-|\bm{B}(\bm{x})|^2)\,d^3\bm{x}.\label{cont_lag}
\end{align}
The second term in the sum is the Lagrangian for a free Maxwell field. The system action is therefore $S=\int L\,dt$, and the general kinetic model must follow from Hamilton's principle applied to $S$. Recall\cite{Landau_1976} that Hamilton's principle entails setting the first variation of the system action equal to zero, $\delta S=0$. At this stage in the discussion, it is not clear how this variation should be performed.

Hamilton's principle can be given a precise meaning as follows. Let $\bm{z}(\bm{z}_o)$ be the Lagrangian configuration map of a given species' phase space fluid. Observe that because the value of the Lagrangian configuration map $\bm{z}(\bm{z}_o)$ gives the phase space location of a particle initially located at $\bm{z}_o$, the Eulerian flow velocity $\bm{V}$ can be expressed in terms of $\bm{z}(\bm{z}_o)$ as
\begin{align}
\frac{d\bm{z}(\bm{z}_o)}{dt}=\bm{V}(\bm{z}(\bm{z}_o)).
\end{align}
Next notice that because particle number is locally conserved, the integral
$
\int_U F(\bm{z})\,d^n\bm{z},
$
where $U\subset M$ moves with the particles, is independent of time. Therefore, the distribution function $F$ can also be expressed in terms of $\bm{z}(\bm{z}_o)$ according to
\begin{align}
F(\bm{z}(\bm{z}_o))\,d^n\bm{z}(\bm{z}_o)=F_o(\bm{z}_o)\,d^n\bm{z}_o,
\end{align}
where $F_o$ is the initial distribution function. It follows that all quantities in the system Lagrangian $L$ can be expressed in terms of the Lagrangian configuration map $\bm{z}(\bm{z}_o)$, the electromagnetic potentials $(\bm{A},\varphi)$, and the initial distribution function $F_o$. Varying the action $S$ can therefore be achieved by varying $\bm{z}(\bm{z}_o)$ and $(\bm{A},\varphi)$.

The Euler-Lagrange equations associated with the variational principle $\delta S=0$ can now be found in principle. Before displaying these equations however, it will be useful to define some linear maps. The first map sends contravariant vector fields on $M$ into the space of covariant vector fields (one-forms) on $M$. This map is encoded in the covariant antisymmetric matrix of functions on $M$
\begin{align}
\omega_{ij}=\partial_j\theta_i-\partial_i\theta_j-\frac{e}{c}\cancel{\epsilon}_{ijk}B^k,\label{cont_symplectic_form}
\end{align}
where $\cancel{\epsilon}_{ijk}=\epsilon_{ijk}$ when $i,j,k\leq 3$ and $\cancel{\epsilon}_{ijk}=0$ otherwise. The second map, which is (minus) the inverse of the first map, sends covariant vector fields on $M$ into the space of contravariant vector fields on $M$. Its  associated antisymmetric contravariant matrix $J^{ij}$ of functions is defined implicitly by the formula
\begin{align}
J^{ij}\omega_{kj}=\delta^i_k.
\end{align}
Note that implicit in the definition of $J^{ij}$ is the assumption that $\omega_{ij}$ gives the components of an invertible matrix. Finally, introduce the linear operators $D_{\bm{B}}\theta_i$, $D_{\bm{E}}K$, and $D_{\bm{B}}K$, where
\begin{align}
D_{\bm{B}}\theta_i[\delta\bm{B}](\bm{z})&=\frac{d}{d\epsilon}\bigg|_0\theta_i(\bm{z};\bm{B}+\epsilon\delta\bm{B})\\
D_{\bm{E}}K[\delta\bm{E}](\bm{z})&=\frac{d}{d\epsilon}\bigg|_0 K(\bm{z};\bm{E}+\epsilon\delta\bm{E},\bm{B})\\
D_{\bm{B}}K[\delta\bm{B}](\bm{z})&=\frac{d}{d\epsilon}\bigg|_0 K(\bm{z};\bm{E},\bm{B}+\epsilon\delta\bm{B}).
\end{align}
The notation here is motivated by the presentation of the Fr\'echet derivative in Ref.\,\onlinecite{Abraham_2008}. Note the distinction between, say, $D_{\bm{E}}K[\delta\bm{E}]$, which is a function on $M$, and $D_{\bm{E}}K[\delta\bm{E}](\bm{z})$ which is a real number. Also note that each of these three operators is a linear map from the space of electric fields $\delta\bm{E}$ or the space of magnetic fields $\delta\bm{B}$ to the space of functions on phase space. 

Now the Euler-Lagrange equations can be written explicitly.
The components of the velocity $\bm{V}$ are given by
\begin{align}
V^i=J^{ij}(\partial_jK-e \cancel{\delta}_j^lE_l+D_{\bm{B}}\theta_j[\partial_t{\bm{B}}]),\label{cont_hamilton_on_M}
\end{align}
where $\cancel{\delta}_j^l=\delta_j^l$ when $j,l\leq 3$ and $\cancel{\delta}_j^l=0$ otherwise. The electromagnetic fields satisfy Maxwell's equations in a polarized, magnetized medium,
\begin{align}
\nabla\times\bm{E}&=-\frac{1}{c}\partial_t\bm{B}\label{cont_faraday}\\
\nabla\times\bm{H}&=\frac{4\pi}{c}\bm{J}_f+\frac{1}{c}\partial_t\bm{D}\\
\nabla\cdot\bm{D}&=4\pi\rho_f\\
\nabla\cdot\bm{B}&=0.\label{cont_divB}
\end{align}
The auxiliary electric and magnetic field are related to the electric and magnetic field according to the constitutive relations
\begin{align}
\bm{D}&=\bm{E}-4\pi\sum D_{\bm{E}}K^{T}[F]\label{cont_D}\\
\bm{H}&=\bm{B}+4\pi\sum (D_{\bm{B}}K^T[F]-D_{\bm{B}}\theta_i^T[V^iF]),\label{cont_H}
\end{align}
where the ``$T$'' superscript denotes the transpose of a linear operator between inner product spaces.\footnote{The inner products on the spaces involved here, i.e. the space of electric fields, the space of magnetic fields, and the space of scalar functions on $M$, are the usual $L^2$ inner products.} Finally, because the local conservation of particles is built explicitly into the variational principle, the preceding equations must also be supplemented by the transport equation
\begin{align}
\partial_tF+\partial_i(FV^i)=0.\label{vlasov}
\end{align}
Equations\,(\ref{cont_hamilton_on_M}-\ref{vlasov}) comprise the general collisionless kinetic plasma model. Assuming $\bm{E}$ can be expressed uniquely in terms of $\bm{D}$ using Eq.\,(\ref{cont_D}), the model prescribes the evolution of the infinite-dimensional state variable $Z=(F,\bm{D},\bm{A})$, where $\bm{A}$ is the vector potential in the $\varphi=0$ gauge.

By employing the method\cite{Burby_thesis_2015} used to identify the gyrokinetic Hamiltonian structure in Ref.\,\onlinecite{Burby_gvm_2015}, the Hamiltonian and Poisson bracket (satisfying the Jacobi identity) for this general kinetic plasma model may each be found. The Hamiltonian functional is given by
\begin{align}
\mathcal{H}(F,\bm{D},\bm{A})&=\sum\int (K-D_{\bm{E}}K[\hat{\bm{E}}])F\,d^n\bm{z}\nonumber\\
&+\frac{1}{8\pi}\int (|\hat{\bm{E}}|^2+|\bm{B}|^2)\,d^3\bm{x},\label{cont_ham}
\end{align}
where $\hat{\bm{E}}$ is the electric field expressed in terms of the auxiliary electric field $\bm{D}$, the magnetic field $\bm{B}$, and the distribution function $F$ using Eq.\,(\ref{cont_D}). Eq.\,(\ref{cont_ham}) expresses the total energy stored in the system as the sum of energy stored in the particles and the energy stored in the electromagnetic field. The Poisson bracket is given by
\begin{align}
\{\mathcal{G},\mathcal{H}\}=&4\pi c\int \bigg(\frac{\delta \mathcal{
G}}{\delta\bm{D}}\cdot\frac{\delta \mathcal{H}}{\delta\bm{A}}-\frac{\delta \mathcal{G} \mathcal{
}}{\delta\bm{A}}\cdot\frac{\delta \mathcal{H}}{\delta\bm{D}}\bigg)\,d^3\bm{x}\nonumber\\
+&\sum\int \bigg(\partial_i\frac{\delta\mathcal{G}}{\delta F}-e\mathcal{E}_i^*\bigg[\frac{\delta\mathcal{G}}{\delta\bm{D}}\bigg]\bigg)J^{ij}\times\nonumber\\
&\qquad\quad\bigg(\partial_j\frac{\delta\mathcal{H}}{\delta F}-e\mathcal{E}_j^*\bigg[\frac{\delta\mathcal{H}}{\delta\bm{D}}\bigg]\bigg)\,F\,d^n\bm{z},\label{cont_poisson}
\end{align}
where the linear operator $\mathcal{E}^*_i$ is given by
\begin{align}
\mathcal{E}^*_i[\delta\bm{D}]=4\pi\cancel{\delta}_i^j\delta D_j+4\pi\frac{c}{e}D_{\bm{B}}\theta_i[\nabla\times\delta\bm{D}].\label{cont_E_star}
\end{align}

This bracket generalizes the brackets given in Refs.\,\onlinecite{Morrison_lifting_2013} and \onlinecite{Burby_gvm_2015}. The bracket in Ref.\,\onlinecite{Morrison_lifting_2013} is recovered as follows. Set $\bm{m}=(\bm{p},w)$, where $\bm{p}\in\mathbb{R}^3$ is the kinetic momentum and $w\in\mathbb{R}^d$ is called the internal degree of freedom\footnote{Note the term ``internal'' is used differently in this paper.} in Ref.\,\onlinecite{Morrison_lifting_2013}. Then set $\theta_i(\bm{z};\bm{B})\dot{z}^i=\bm{p}\cdot\dot{\bm{x}}+\vartheta(w)\cdot\dot{w}$, where $\vartheta$ is a $d$-component vector that only depends on $w$.  Equation \,(\ref{cont_poisson}) then recovers the bracket in Ref.\,\onlinecite{Morrison_lifting_2013}. The bracket in Ref.\,\onlinecite{Burby_gvm_2015} is recovered by Eq.\,(\ref{cont_poisson}) when $\bm{m}=v_\parallel$ and $\theta_i$ has no $\bm{B}$-dependence.

%%%%%%%%%%%%%%%%%%%%%%%%%%%%%%%%%%%%%%%%%%%%%%%%%%%%%%%%%%%%%%%%%%%
\section{Finite dimensional truncation of the continuum model\label{finite}}
In this section I will show that any continuum kinetic model that is a special case of the model in Section \ref{continuum} has a continuous-time Hamiltonian discretization. Because all known fully-electromagnetic Hamiltonian kinetic models are special cases of the model in Section \ref{continuum}, this section will give a procedure for discretizing a very broad class of kinetic models while preserving Hamiltonian structure.
This procedure represents a significant generalization of similar procedures used by other authors in more specialized scenarios. Hamiltonian Continuous-time discretizations of the Vlasov-Maxwell system were developed in Refs.\,\onlinecite{Shadwick_2014}, \onlinecite{Qin_nuc_2016}, \onlinecite{Xiao_2015}, \onlinecite{He_2016}, and \onlinecite{Kraus_2016_arXiv}. A Hamiltonian continuous-time discretization of the low-flow drift-kinetic Vlasov-Maxwell system was found in in Ref.\,\onlinecite{Evstatiev_2014}. 

%This section will develop a finite-dimensional approximation to the general, infinite-dimensional kinetic model presented in the previous section. Passing from infinitely many dimensions to a finite number of them will be accomplished by discretizing the electromagnetic fields and the distribution function. Time will remain a continuous variable in this process. By performing the discretization at the level of the Lagrangian given in Eq.\,(\ref{cont_lag}), the finite-dimensional model will inherit its parent model's Hamiltonian structure.
\subsection{Formulation of the finite-dimensional model}
At the continuum level, the electromagnetic fields and potentials are naturally elements of different function spaces. The potentials $(\bm{A},\varphi)$ must each be square integrable in order to ensure the minimal coupling integral in $L$, $\int (\bm{J}_f\cdot\bm{A}-\rho_f\varphi)\,d^3\bm{x}$, is finite. The gradient of the scalar potential and the curl of the vector potential must also each be square integrable in order to ensure that the energy stored in the electric and magnetic fields are finite. Thus, $\varphi\in H^1(\text{grad})$ and $\bm{A}\in H^1(\text{curl})$, where
\begin{align}
H^1(\text{grad})	&=\{\psi\in L^2(Q)\mid \nabla\psi\in \bm{L}^2(Q) \}\\
H^1(\text{curl})&=\{\bm{\alpha}\in \bm{L}^2(Q)\mid \nabla\times\bm{\alpha}\in \bm{L}^2(Q)\}.
\end{align}
Here $L^2(Q)$ and $\bm{L}^2(Q)$ are the square-integrable scalar fields and vector fields defined on the spatial domain $Q$. It follows that $\bm{E}=-c^{-1}\partial_t\bm{A}-\nabla\varphi$ and $\bm{B}=\nabla\times\bm{A}$ are elements of the spaces $H^1(\text{curl})$ and $H^1(\text{div})$, respectively,\footnote{The curl of $\bm{E}$ is square integrable because $\nabla\times\bm{E}=-c^{-1}\partial_t\nabla\times \bm{A}$ and $\nabla\times\bm{A}$ is square integrable. The divergence of $\bm{B}$ is square integrable because $\nabla\cdot\bm{B}=\nabla\cdot\nabla\times\bm{A}=0$.} where
\begin{align}
H^1(\text{div})=\{\bm{\beta}\in\bm{L}^2(Q)\mid \nabla\cdot\bm{\beta}\in L^2(Q)\}.	
\end{align}
Note that because $\nabla\times\nabla\psi=\nabla\cdot\nabla\times\bm{\alpha}=0$ the differential operators of vector calculus are well behaved with respect to these spaces in the sense that
\begin{align}
\nabla(H^1(\text{grad}))&\subset H^1(\text{curl})\\
\nabla\times(H^1(\text{curl}))&\subset H^1(\text{div})\\
\nabla\cdot(H^1(\text{div}))&\subset L^2(Q).	
\end{align}

In order to discretize the electromagnetic fields and potentials, the spaces $H^1(\text{grad}),H^1(\text{curl}),H^1(\text{div}),$ and $L^2(Q)$ will be replaced with finite dimensional subspaces
\begin{align}
W_0&\subset H^1(\text{grad})\\
W_1&\subset H^1(\text{curl})\\
W_2&\subset H^1(\text{div})\\
W_3&\subset L^2(Q).	
\end{align}
The discrete potentials $(\mathbb{A},\Phi)$ and fields $(\mathbb{E},\mathbb{B})$ will satisfy $\Phi\in W_0$, $\mathbb{A}\in W_1$, $\mathbb{E}\in W_1$, and $\mathbb{B}\in W_2$. Moreover, in order to ensure that discrete versions of the gradient, curl, and divergence operators respect the identities $\nabla\times\nabla\psi=0$ and $\nabla\cdot\nabla\times\bm{\alpha}=0$, the subspaces will be chosen to satisfy
\begin{align}
\nabla(W_0)&\subset W_1\label{conform_grad}\\
\nabla\times(W_1)&\subset W_2\\
\nabla\cdot(W_2)&\subset W_3.\label{conform_div}	
\end{align}
The discrete gradient, curl, and divergence operators, $\mathsf{G},\mathsf{C}$ and $\mathsf{D}$, are then defined by restricting their continuous counterparts to the subspaces $W_0,W_1,$ and $W_2$, respectively. For instance, if $\psi\in W_0$, then $\G\psi=\nabla\psi\in W_1$. There are many possible choices for the subspaces $W_i$. Truncated Fourier bases are perhaps the simplest options. More generally, any spaces that fit into the framework of finite element exterior calculus\cite{Arnold_2006_acta} (FEEC) would work. Note that FEEC discretizations were used in Vlasov-Maxwell simulations in Refs.\,\onlinecite{He_2016} and \onlinecite{Kraus_2016_arXiv}.

In order to discretize the distribution function $F$, macroparticles\cite{Shadwick_2014} will be introduced. This entails expressing $F$ as
\begin{align}
F(\bm{x},\bm{m})=\sum_{\text{p}}\delta^m(\bm{m}-\bm{m}_{\text{p}})\mathcal{W}(\bm{x}-\bm{x}_{\text{p}}),	
\end{align}
where $\text{p}$ indexes a collection of $N$ macroparticles. Each macroparticle is assigned a location $\bm{z}_{\text{p}}=(\bm{x}_{\text{p}},\bm{m}_{\text{p}})$ in the single-particle phase space $M$ and a finite spatial extent given by the support of the differentiable weight function $\mathcal{W}$. The weight function has compact support and satisfies $\int \mathcal{W}(\bm{x})\,d^3\bm{x}=1$. Instead of prescribing an evolution law for $F$, the finite-dimensional kinetic model will prescribe an evolution law for the macroparticles $\bm{z}_{\p}$.

Because macroparticles are not point-like, the Lagrangian for an individual macroparticle is different from the Lagrangian $\ell$ for point-like particles. The single macroparticle Lagrangian $\langle\ell_\p\rangle$ is given by convolving the point-particle Lagrangian with the weight function $\mathcal{W}$,
\begin{align}
\langle\ell_\p\rangle &=\int \theta_{\p i}(\bm{z};\mathbb{B})\dot{z}_\p^i	\delta^m({\bm{m}-\bm{m}_\p})\mathcal{W}(\bm{x}-\bm{x}_\p)\,d^n\bm{z}\nonumber\\
&-\int K_\p(\bm{z};\mathbb{E},\mathbb{B})\delta^m({\bm{m}-\bm{m}_\p})\mathcal{W}(\bm{x}-\bm{x}_\p)\,d^n\bm{z}\nonumber\\
&+\frac{e}{c}\int (\mathbb{A}(\bm{x})\cdot\dot{\bm{x}}_\p-c\,\Phi(\bm{x}))\mathcal{W}(\bm{x}-\bm{x}_\p)\,d^3\bm{x}.
\end{align}
Here the discrete electric and magnetic fields are given by $\mathbb{E}=-c^{-1}\partial_t\mathbb{A}-\G\Phi$ and $\mathbb{B}=\C \mathbb{A}$. Note that by introducing the quantities
\begin{align*}
\langle\theta_\p\rangle_i(\bm{z};\mathbb{B}) &=\int \theta_{\p i}(\bar{\bm{z}};\mathbb{B})\delta^m(\bar{\bm{m}}-\bm{m})\mathcal{W}(\bar{\bm{x}}-\bm{x})\,d^n\bar{\bm{z}}\\
\langle K_\p\rangle(\bm{z};\mathbb{E},\mathbb{B})&=\int K_\p(\bar{\bm{z}};\mathbb{E},\mathbb{B})\delta^m(\bar{\bm{m}}-\bm{m})\mathcal{W}(\bar{\bm{x}}-\bm{x})\,d^n\bar{\bm{z}}\\
\langle\mathbb{A}\rangle(\bm{x})&=	\int \mathbb{A}(\barx)\mathcal{W}(\barx-\bm{x})\,d^3\barx\\
\langle\Phi\rangle(\bm{x})&=\int \Phi(\barx)\mathcal{W}(\barx-\bm{x})\,d^3\barx,
\end{align*}
the macroparticle Lagrangian can be written
\begin{align}
\lp=&\langle\theta_\p\rangle_i(\bm{z}_\p;\mathbb{B})\dot{z}_\p^i-\langle K_\p\rangle(\bm{z}_\p;\mathbb{E},\mathbb{B})\nonumber\\
&+\frac{e_\p}{c}\langle\mathbb{A}\rangle(\bm{x}_\p)\cdot\dot{\bm{x}}_\p-e_\p\langle\Phi\rangle(\bm{x}_\p),\label{lp_nice}
\end{align}
which is formally very similar to $\ell$ in Eq.\,(\ref{sin_part}). The total Lagrangian for all macroparticles, $\mathbb{L}_\chi$, is therefore given by summing $\lp$ over all particles,
\begin{align}
\mathbb{L}_\chi &=\sum_\p\lp\nonumber.	
\end{align}
Likewise, the Lagrangian for the entire discrete field-particle system is given by
\begin{align}
\mathbb{L}=\mathbb{L}_\chi+\frac{1}{8\pi}\int (|\mathbb{E}|^2-|\mathbb{B}|^2)\,d^3\bm{x}.	
\end{align}

A finite-dimensional approximation to the general kinetic plasma model presented in the previous section may now be defined by applying Hamilton's principle to the Lagrangian $\mathbb{L}$. The discrete action is $\mathbb{S}=\int \mathbb{L}\,dt$. The quantities that must be varied when varying the discrete action are each of the macroparticle trajectories, $\bm{z}_\p(t)$, and the discrete electromagnetic potentials $(\mathbb{A}(t),\Phi(t))$. 

Before displaying the Euler-Lagrange equations associated with the discrete version of Hamilton's principle, it is again useful to define some linear maps. The first linear map sends contravariant vector fields on $M$ to covariant vector fields on $M$. It is encoded in the antisymmetric matrix of components
\begin{align}
\langle\omega_\p\rangle_{ij}&=\partial_j\langle\theta_\p\rangle_i-\partial_i\langle\theta_\p\rangle_j\nonumber\\
&-\frac{e_\p}{c}\cancel{\epsilon}_{ijk}\langle\mathbb{B}\rangle^k,\label{disc_symp}
\end{align}
where 
\begin{align}
\langle\mathbb{B}\rangle(\bm{x})=\int \mathbb{B}(\bar{\bm{x}})\mathcal{W}(\bar{\bm{x}}-\bm{x})\,d^3\bar{\bm{x}}.
\end{align}
The second map is (minus) the inverse of the first. Its associated antisymmetric matrix of functions $\langle J_\p\rangle^{ij}$ is defined implicitly by the formula
\begin{align}
\langle J_\p\rangle^{ij}\langle\omega_\p\rangle_{kj}=\delta^i_k.
\end{align}
Finally, introduce the linear maps $D_{\mathbb{B}}\langle\theta_\p\rangle_i$, $D_{\mathbb{E}}\langle K_\p\rangle$, and $D_{\mathbb{B}}\langle K_\p\rangle$, where
\begin{align}
D_{\mathbb{B}}\langle\theta_\p\rangle_i[\delta\mathbb{B}](\bm{z})&=\frac{d}{d\epsilon}\bigg|_0\langle\theta_\p\rangle_i(\bm{z};\mathbb{B}+\epsilon\delta\mathbb{B})\\
D_{\mathbb{E}}\langle K_\p\rangle[\delta\mathbb{E}](\bm{z})&=\frac{d}{d\epsilon}\bigg|_0\langle K_\p\rangle(\bm{z};\mathbb{E}+\epsilon\delta\mathbb{E},\mathbb{B})\\
D_{\mathbb{B}}\langle K_\p\rangle[\delta\mathbb{B}](\bm{z})&=\frac{d}{d\epsilon}\bigg|_0\langle K_\p\rangle(\bm{z};\mathbb{E},\mathbb{B}+\epsilon\delta\mathbb{B}).
\end{align}
Note that $D_{\mathbb{B}}\langle\theta_\p\rangle_i$ and $D_{\mathbb{B}}\langle K_\p\rangle$ map the space $W_2$ into the space of functions on $M$. Similarly, $D_{\mathbb{E}}\langle K_\p\rangle$ maps $W_1$ into the space of functions on $M$.

Now the Euler-Lagrange equations associated with the discrete action $\mathbb{S}$ can be written explicitly. The components of the phase-space velocity of macroparticle $\p$ are given by
\begin{align}\label{disc_par}
\dot{z}^i_\p=\langle J_\p\rangle^{ij}\Big(\partial_j \langle K_\p\rangle-e_\p \cancel{\delta}^l_j\langle \mathbb{E}\rangle_l+D_{\mathbb{B}}\langle\theta_\p\rangle_j[\partial_t\mathbb{B}]\Big),
\end{align}
where
\begin{align}
\langle\mathbb{E}\rangle(\bm{x})=\int\mathbb{E}(\barx)\mathcal{W}(\barx-\bm{x})\,d^3\barx,
\end{align}
and all scalar functions on $M$ appearing on the right hand side of Eq.\,(\ref{disc_par}) are evaluated at $\bm{z}_\p$. The discrete electromagnetic fields satisfy
\begin{align}
\C\mathbb{E}&=-\frac{1}{c}\partial_t\mathbb{B}\label{disc_faraday}\\
\C^T\mathbb{H}&=\frac{4\pi}{c}\Pi_1\sum_\p\dot{\bm{x}}_\p\mathcal{W}_{\bm{x}_\p}+\frac{1}{c}\partial_t\mathbb{D}\label{disc_ampere}\\
\G^T\mathbb{D}&=-4\pi\Pi_0\sum_\p\mathcal{W}_{\bm{x}_\p}\label{disc_gauss}\\
\D\mathbb{B}&=0,\label{disc_divB}
\end{align}
which represent the discrete analogues of Maxwell's equations in a poralized and magnetized medium. 	Here $\mathcal{W}_{\bm{x}_\p}(\bm{x})=\mathcal{W}(\bm{x}-\bm{x}_\p)$ and $\Pi_0,\Pi_1$ are the orthogonal projections onto the spaces $W_0$ and $W_1$, respectively. The discrete auxiliary electric and magnetic field are related to the discrete electric and magnetic field according to the constitutive relations
\begin{align}
\mathbb{D}&=\mathbb{E}-4\pi\sum_\p D_{\mathbb{E}}\langle K_\p\rangle^T[\delta^n_{\bm{z}_\p}]\label{disc_D}\\
\mathbb{H}&=\mathbb{B}+4\pi\sum_\p\Big(D_{\mathbb{B}}\langle K_\p\rangle^T[\delta^n_{\bm{z}_\p}]-D_{\mathbb{B}}\langle\theta_\p\rangle_i^T[\dot{z}_\p^i\delta^n_{\bm{z}_\p}]\Big),\label{disc_H}
\end{align}
where $\delta^n_{\bm{z}_\p}(\bm{z})=\delta^n(\bm{z}-\bm{z}_\p)$. Equations\,(\ref{disc_par}-\ref{disc_H}) comprise the finite-dimensional collisionless kinetic plasma model, which are also written explicitly in component form in the Appendix. Note that because $\G(W_0)\subset W_1$ and $\C(W_1)\subset W_2$, the discrete Gauss' Law (\ref{disc_gauss}) and the discrete $\text{div}(\bm{B})$-constraint (\ref{disc_divB}) are merely constraints on the initial conditions for $\bb{E}$ and $\bb{B}$.

The Hamiltonian structure underlying this model can be found using essentially the same procedure used earlier to find the Hamiltonian structure of collisionless kinetic theory. In fact, the procedure in finite dimensions is even more simple than it is in the continuum theory. It begins by recognizing that there is an alternative Lagrangian for the finite-dimensional model. The alternative Lagrangian is
\begin{align}
\mathfrak{L}=&\sum_\p \langle\theta_\p\rangle_i(\bm{z}_\p;\mathbb{B})\dot{z}_\p^i+\sum_\p\frac{e_\p}{c}\langle\mathbb{A}\rangle(\bm{x}_\p)\cdot\dot{\bm{x}}_\p\nonumber\\
&-\frac{1}{4\pi c}\int \mathbb{D}(\bm{x})\cdot\partial_t\mathbb{A}(\bm{x})\,d^3\bm{x}-\mathfrak{H}(\bar{\bm{z}},\mathbb{D},\mathbb{A}),\label{alt_lag}
\end{align}
where $\bar{\bm{z}}=(\bm{z}_1,\bm{z}_2,\dots,\bm{z}_N)$ is the collection macroparticles, and $\mathfrak{H}$ is given by
\begin{align}
\mathfrak{H}(\bar{\bm{z}},\mathbb{D},\mathbb{A})&=\sum_\p\Big(\langle K_\p\rangle(\bm{z}_\p)-D_{\mathbb{E}}\langle K_\p\rangle[\hat{\mathbb{E}}](\bm{z}_\p)\Big)\nonumber\\
&+\frac{1}{8\pi}\int \Big(|\hat{\mathbb{E}}|^2+|\mathbb{B}|^2\Big)\,d^3\bm{x}.\label{disc_ham}
\end{align}
Here $\hat{\mathbb{E}}$ is the discrete electric field expressed in terms of $\bar{\bm{z}},\mathbb{D},\mathbb{B}$ using the constitutive relation\,(\ref{disc_D}). When varying the action $\mathfrak{S}=\int \mathfrak{L}\,dt$ associated with this Lagrangian, the quantities that are varied are $Z=(\bar{\bm{z}},\mathbb{D},\mathbb{A})$. It is straightforward to verify that the Euler-Lagrange equations associated with $\mathfrak{S}$ are equivalent to Eqs.\,(\ref{disc_par}-\ref{disc_H}). Next, notice that $\mathfrak{L}$ has the general form
\begin{align}
\mathfrak{L}=\Theta(Z)_\alpha\dot{Z}^\alpha-\mathfrak{H}(Z),\label{phase_space_lag}
\end{align} 
where $\dot{Z}^\alpha$ is the $\alpha$'th component of the vector $\dot{Z}=(\dot{\bar{\bm{z}}},\partial_t\mathbb{D},\partial_t\mathbb{A})$ and the $\Theta_\alpha$ are $N$ functions of $Z$ that can be read off of Eq.\,(\ref{alt_lag}). In other words, $\mathfrak{L}$ is a phase space Lagrangian. It follows from the general theory of phase space Lagrangians\cite{Cary_1983} that the Poisson bracket for this finite-dimensional system is given by
\begin{align}
\{\mathfrak{G},\mathfrak{H}\}=(\partial_\alpha\mathfrak{G})\mathfrak{J}^{\alpha\beta}(\partial_\beta\mathfrak{H}),\label{gen_poisson_formula}
\end{align}
where $\mathfrak{G},\mathfrak{H}$ are arbitrary functions of $Z$, and $\mathfrak{J}^{\alpha\beta}$ is (minus) the inverse of the matrix
\begin{align}
\Omega_{\alpha\beta}=\partial_\beta\Theta_\alpha-\partial_\alpha\Theta_\beta,
\end{align}
and the Hamiltonian is given by $\mathfrak{H}$.

Applying the formula\,(\ref{gen_poisson_formula}) eventually produces the expression
\begin{align}\label{disc_poisson}
\{\mathfrak{G},\mathfrak{H}\}=&4\pi c\int\left(\frac{\delta\mathfrak{G}}{\delta\mathbb{D}}\cdot\frac{\delta\mathfrak{H}}{\delta\mathbb{A}}-\frac{\delta\mathfrak{H}}{\delta\mathbb{D}}\cdot\frac{\delta\mathfrak{G}}{\delta\mathbb{A}}\right)\,d^3\bm{x}\nonumber\\
&+\sum_\p\bigg(\frac{\partial\mathfrak{G}}{\partial z_{\p}^i}-e_\p\langle\mathcal{E}_\p^*\rangle_i\left[\frac{\delta\mathfrak{G}}{\delta \mathbb{D}}\right](\bm{z}_\p)\bigg)\langle J_\p\rangle^{ij}\nonumber\\
&\qquad\cdot\bigg(\frac{\partial\mathfrak{H}}{\partial z_{\p}^j}-e_\p\langle\mathcal{E}_\p^*\rangle_j\left[\frac{\delta\mathfrak{H}}{\delta \mathbb{D}}\right](\bm{z}_\p)\bigg)
\end{align}
for the finite-dimensional model's Poisson bracket. Here the operator $\langle\mathcal{E}^*_\p\rangle_i$ is given by
\begin{align}
\langle\mathcal{E}^*_\p\rangle_i[\delta\mathbb{D}]=4\pi \cancel{\delta}_i^j\langle\delta\mathbb{D}\rangle_j+4\pi \frac{c}{e_\p}D_{\mathbb{B}}\langle\theta_\p\rangle_i[\C\delta\mathbb{D}],\label{disc_estar}
\end{align}
where $\langle\delta\mathbb{D}\rangle(\bm{x})=\int \delta\mathbb{D}(\barx)\mathcal{W}(\barx-\bm{x})\,d^3\barx$. Together with the expression\,(\ref{disc_ham}), Eq.\,(\ref{disc_poisson}) completely specifies the Hamiltonian structure of the finite-dimensional collisionless kinetic model derived in this section. The component-based form of the Poisson bracket is given in the Appendix.

%\subsection{Gauge invariance in finite dimensions}

%%%%%%%%%%%%%%%%%%%%%%%%%%%%%%%%%%%%%%%%%%%%%%%%%%%%%%%%%%%%%%%%%%%
\section{Example: relativistic Vlasov-Maxwell with spin\label{rel_spin}}
As a first concrete example of the theory developed in Sections \ref{continuum} and \ref{finite}, this section will apply the theory to the relativistic Vlasov-Maxwell system with spin. This system couples Maxwell's equations to an ensemble of micro-localized wave packet solutions of the Dirac equation. Because the wave packets are localized in both Fourier space and real space, they obey classical equations of motion. Moreover, due to the spin, each wave packet acquires electric and magnetic dipole moments. While a nonrelativistic limit of this system was studied in Ref.\,\onlinecite{Brodin_2008}, the full relativistic model has not been studied previously.

%First the continuum model will be presented. As will become explicit shortly, the entire model is encoded in the single-particle Lagrangian $\ell$. Then the continuous-time discrtization of the continuum model will be presented. Much as in the continuum theory, the ``semi-discrete'' model is encoded in the single-macroparticle Lagrangian $\langle\ell_\p\rangle$.

The classical phase space $M$ for a spin-$1/2$ particle is given in terms of the notation in Section \ref{continuum} as
\begin{align}
\bm{m}=(\bm{p},\mathsf{Z}),
\end{align}
where $\bm{p}\in\mathbb{R}^3$ is the particle kinetic momentum and $\mathsf{Z}\in\mathbb{C}^2$ is a (dimensionless) two-component spinor representing the polarization of the (positive-energy) Dirac field. The dimension of the internal state space for a classical particle with spin $1/2$ is therefore $m=3+4=7$, while the dimension of $M$ is $n=3+7=10$. Associated with the spinor $\mathsf{Z}$ is the so-called spin vector $\bm{S}=\hbar \mathsf{Z}^\dagger\bm{\sigma}\mathsf{Z}/2\in \mathbb{R}^3$. Here $\bm{\sigma}=\sigma_x \bm{e}_x+\sigma_y \bm{e}_y+\sigma_z \bm{e}_z$ is a three-component vector with complex $2\times 2$ matrix components $\sigma_i$. The vectors $\bm{e}_i$ are the standard basis vectors in $\mathbb{R}^3$ and the matrices $\sigma_i$ are the well-known Pauli matrices.   

The Lagrangian $\ell$ governing the classical dynamics of a spin-$1/2$ particle was derived recently in Ref.\,\onlinecite{Ruiz_2015_pla}. By substituting an appropriate wave-packet ansatz into the Lagrangian for a Dirac field in prescribed electromagnetic fields, those authors found that $\ell$ is given by Eq.\,(\ref{sin_part}) with
\begin{align}
\theta_i(\bm{z};\bm{B})\dot{z}^i&=\bm{p}\cdot\dot{\bm{x}}+\frac{i\hbar}{2}\Big(\mathsf{Z}^\dagger\dot{\mathsf{Z}}-\dot{\mathsf{Z}}^\dagger\mathsf{Z}\Big)\label{vm_theta}\\
K(\bm{z};\bm{E},\bm{B})&=\gamma m c^2+\bm{S}\cdot \bm{\Omega},
\end{align}
where
\begin{align}
\gamma&=\sqrt{1+\frac{|\bm{p}|^2}{m^2c^2}}=\frac{1}{\sqrt{1-|\bm{\beta}|^2}}\\
\bm{\beta}&=\frac{\bm{p}}{\gamma m c}\\
\bm{\Omega}&=\Upsilon_{\bm{B}}\cdot\bm{B}+\Upsilon_{\bm{E}}\cdot\bm{E}.
\end{align}
The $\bm{p}$-dependent tensors $\Upsilon_{\bm{B}},\Upsilon_{\bm{E}}$ are given by
\begin{align}
\Upsilon_{\bm{B}}&=\frac{e}{mc}\left[\left(\frac{g}{2}-1+\frac{1}{\gamma}\right)\mathbb{I}-\left(\frac{g}{2}-1\right)\frac{\gamma}{\gamma+1}\bm{\beta}\bm{\beta}\right]\\
\Upsilon_{\bm{E}}&=\frac{e}{mc}\left[\left(\frac{g}{2}-\frac{\gamma}{\gamma+1}\right)\bm{e}_i\bm{e}_j\epsilon_{ijk}\beta_k\right].
\end{align}
Here $g$ is the so-caled $g$-factor, which quantifies a particle's anomolous magnetic moment. For electrons $g\approx 2.0$, while for protons $g\approx5.5$. The Euler-Lagrange equations associated with this Lagrangian reproduce the famous Bargmann-Michel-Telegdi (BMT) equations.\cite{Bargmann_1959}

The relativistic Vlasov-Maxwell system with spin springs forth from $\ell$ as in Section \ref{continuum}. According to Eqs.\,(\ref{cont_D}-\ref{cont_H}), the consitutive relations defining the auxiliary electric field and auxiliary magnetic field are given by
\begin{align}
\bm{D}&=\bm{E}-4\pi\sum \int\bm{S}\cdot\Upsilon_{\bm{E}}\,F d^7\bm{m}.\nonumber\\
\bm{H}&=\bm{B}+4\pi\sum \int\bm{S}\cdot\Upsilon_{\bm{B}}\,F\,d^7\bm{m}
\end{align}
Thus, the plasma behaves as a polarized and magnetized medium with polarization and magnetization densities given by
\begin{align}
\bm{P}&=-\sum\int\bm{S}\cdot\Upsilon_{\bm{E}}\,F\,d^7\bm{m}\\
\bm{M}&=-\sum\int\bm{S}\cdot\Upsilon_{\bm{B}}\,F\,d^7\bm{m}.
\end{align}
Observe that the polarization density vanishes is the non-relativistic limit, while the magnetization density does not. The evolution of the electromagnetic field is then determined by substituting these constitutive relations into Maxwell's equations in a medium, Eqs.\,(\ref{cont_faraday}-\ref{cont_divB}). Finally, the evolution of the distribution function is given by the Vlasov equation (\ref{vlasov}), where the Eulerian velocity in phase space $\bm{V}$ is determined as follows. Decompose the $\bm{m}$-component of $\bm{V}$ as $\bm{w}=(\bm{f},\nu)$, where $\bm{f}$ gives a particle's $\bm{p}$-velocity (i.e. force) and $\nu$ gives a particle's $\mathsf{Z}$-velocity. The components of $\bm{V}$ are then given by Eq.\,(\ref{cont_hamilton_on_M}), which becomes
\begin{align}
\bm{u}&=\frac{\delta K}{\delta\bm{p}}\\
\bm{f}&=-\frac{\delta K}{\delta\bm{x}}+e\bm{E}+\frac{e}{c}\frac{\delta K}{\delta\bm{p}}\times\bm{B}\\
\nu&=-\frac{i}{2\hbar}\frac{\delta K}{\delta \mathsf{Z}}.
\end{align}
Here the finite-dimensional functional derivatives of a scalar function $\chi$ on $M$ are defined by 
\begin{align}
\frac{d}{d\epsilon}\bigg|_0\chi(\bm{z}+\epsilon \delta\bm{z})=\frac{\delta\chi}{\delta\bm{x}}\cdot\delta\bm{x}+\frac{\delta\chi}{\delta\bm{p}}\cdot\delta\bm{p}+\text{Re}\left(\frac{\delta\chi}{\delta\mathsf{Z}}^\dagger\delta\mathsf{Z}\right).
\end{align}
The derivatives of $K$ can be computed, giving
\begin{align}
\frac{\delta K}{\delta\bm{x}}&=\frac{\delta}{\delta\bm{x}}(\bm{S}\cdot\bm{\Omega})\\
\frac{\delta K}{\delta \bm{p}}&=\frac{\bm{p}}{\gamma m}+\frac{\delta}{\delta\bm{p}}(\bm{S}\cdot\bm{\Omega})\\
\frac{\delta K}{\delta\mathsf{Z}}&=\hbar \,(\bm{\sigma}\cdot\bm{\Omega})\mathsf{Z}.
\end{align}
In the non-relativistic limit, $\bm{S}\cdot\bm{\Omega}$ is independent of $\bm{p}$. Thus, the above expressions show a particle's $\bm{x}$-velocity $\bm{u}$ is parallel to its kinetic momentum $\bm{p}$ in the nonrelativistic limit. In contrast, with relativistic effects fully included, $\bm{p}$ is no longer parallel to $\bm{u}$.

The relativistic Vlasov-Maxwell system with spin is an infinite-dimensional Hamiltonian system with the following Hamiltonian structure. The Hamiltonian is given by substituting the definitions of $K$ and $\bm{D}$ into Eq.\,(\ref{cont_ham}), leading to
\begin{align}
&\mathcal{H}(F,\bm{D},\bm{A})=\sum\int (\gamma m c^2+\bm{S}\cdot\Upsilon_{\bm{B}}\cdot\bm{B})\,F\,d^n\bm{z}\nonumber\\
&+\frac{1}{8\pi}\int \left|\bm{D}+4\pi\sum \int\bm{S}\cdot\Upsilon_{\bm{E}}\,F d^7\bm{m}\right|^2\,d^3\bm{x}\nonumber\\
&\hspace{6em}+\frac{1}{8\pi}\int |\bm{B}|^2\,d^3\bm{x}.
\end{align}
The Poisson bracket is given by Eq.\,(\ref{cont_poisson}), where the two model-dependent parameters, $J^{ij}$ and $\mathcal{E}_i^*$, are given by
\begin{align}
(\partial_if)J^{ij}(\partial_j g)&=\frac{\delta f}{\delta\bm{x}}\cdot\frac{\delta g}{\delta\bm{p}}-\frac{\delta f}{\delta\bm{p}}\cdot\frac{\delta g}{\delta\bm{x}}\nonumber\\
&+\frac{e}{c}\bm{B}\cdot\frac{\delta f}{\delta\bm{p}}\times\frac{\delta g}{\delta\bm{p}}+\frac{1}{2\hbar}\text{Im}\frac{\delta f}{\delta\mathsf{Z}}^{\dagger}\frac{\delta g}{\delta\mathsf{Z}}\\
\delta{z}^i\mathcal{E}_i^*[\delta\bm{D}]&=4\pi\delta{\bm{x}}\cdot \delta\bm{D}.
\end{align}
The explicit form of the bracket is therefore
\begin{widetext}
\begin{align}
&\{\mathcal{G},\mathcal{H}\}=\sum\int \bigg[\px\left(\func{\calG}{F}\right)\cdot\pp\left(\func{\calH}{F}\right)-\pp\left(\func{\calG}{F}\right)\cdot\px\left(\func{\calH}{F}\right)+\frac{e}{c}\bm{B}\cdot \pp\left(\func{\calG}{F}\right)\times\pp\left(\func{\calH}{F}\right)\bigg]\,F\,d^n\bm{z}\nonumber\\
&+\sum\int \frac{1}{2\hbar}\text{Im}\left[\frac{\delta }{\delta\mathsf{Z}}\left(\func{\calG}{F}\right)^\dagger\frac{\delta}{\delta\mathsf{Z}}\left(\func{\calH}{F}\right)\right]\,F\,d^n\bm{z}+\sum\int4\pi e\bigg[\pp\left(\func{\calG}{F}\right)\cdot\func{\calH}{\bm{D}}-\func{\calG}{\bm{D}}\cdot\pp\left(\func{\calH}{F}\right)\bigg]\,F\,d^n\bm{z}\nonumber\\
&\hspace{15em}+\int 4\pi c \bigg[\func{\calG}{\bm{D}}\cdot\func{\calH}{\bm{A}}-\func{\calG}{\bm{A}}\cdot\func{\calH}{\bm{D}}\bigg]\,d^3\bm{x}.
\end{align}
\end{widetext}
Aside from the fact that this bracket uses $\mathsf{Z}$ instead of $\bm{S}$ to encode the spin degrees of freedom, it agrees in an essential way with the bracket for non-relativistic Vlasov-Maxwell with spin given in Ref.\,\onlinecite{Morrison_lifting_2013}. The bracket in Ref.\,\onlinecite{Morrison_lifting_2013} is obtained from the above bracket by restricting to distribution functions $F$ that depend on $\mathsf{Z}$ only via $\bm{S}$.
%Note that in this example the polarization and magnetization densities are completely determined by the fuction $K$ while the Poisson bracket is completely determined by the set of functions $\theta_i$. When $\theta_i$ has a non-trivial dependence on $\bm{B}$ this will no longer be true.

As I have shown to be true more generally in Section \ref{continuum}, the relativistic Vlasov-Maxwell system with spin can be discretized in space and particles while preserving its Hamiltonian structure. I will demonstrate this fact explicitly by following the prescription in Section \ref{finite}. Much as in the continuum theory, the development begins with the single-macroparticle Lagrangian $\langle\ell_\p\rangle$ given in Eq.\,(\ref{lp_nice}), with
\begin{align}
\langle\theta_\p\rangle_i(\bm{z};\mathbb{B})\dot{z}^i&=\bm{p}\cdot\dot{\bm{x}}+\frac{i\hbar}{2}\Big(\mathsf{Z}^\dagger\dot{\mathsf{Z}}-\dot{\mathsf{Z}}^\dagger\mathsf{Z}\Big)\label{disc_svm_theta}\\
\langle K_\p\rangle(\bm{z};\mathbb{E},\mathbb{B})&=\gamma_\p m_\p c^2+\bm{S}\cdot\langle\bm{\Omega}_\p\rangle,\label{disc_svm_K}
\end{align}
where
\begin{align}
\langle\bm{\Omega}_\p\rangle=\Upsilon_{\bm{B}\p}\cdot\langle\mathbb{B}\rangle+\Upsilon_{\bm{E}\p}\cdot\langle\mathbb{E}\rangle.
\end{align}
Note that the subscript $\p$ is being affixed to quantities that may vary from macroparticle to macroparticle. The discrete auxiliary electromagnetic fields are then found by substituting Eqs.\,(\ref{disc_svm_theta}-\ref{disc_svm_K}) into Eqs.\,(\ref{disc_D}-\ref{disc_H}), leading to
\begin{align}
\mathbb{D}&=\mathbb{E}-4\pi\sum_\p \Pi_1\Big(\bm{S}_\p\cdot\Upsilon_{\bm{E}\p}\mathcal{W}_{\bm{x}_\p}\Big)\label{rel_vm_D}\\
\mathbb{H}&=\mathbb{B}+4\pi\sum_\p\Pi_2\Big(\bm{S}_\p\cdot\Upsilon_{\bm{B}\p}\mathcal{W}_{\bm{x}_\p}\Big),\label{rel_vm_H}
\end{align}
where $\Pi_1$ and $\Pi_2$ are the orthgonal projections onto $W_1$ and $W_2$, respectively. The evolution of the discrete electromagnetic field is therefore given by substituting Eqs.\,(\ref{rel_vm_D}-\ref{rel_vm_H}) into the discrete Maxwell equations in a medium, Eqs.\,(\ref{disc_faraday}-\ref{disc_divB}). Finally, the macroparticle evolution equation is given by Eq.\,(\ref{disc_par}), which becomes
\begin{align}
\dot{\bm{x}}_\p&=\frac{\bm{p}_\p}{\gamma_\p m_\p}+\frac{\delta}{\delta\bm{p}_\p}(\bm{S}_\p\cdot\langle\bm{\Omega}_\p\rangle)\\
\dot{\bm{p}}_\p&=-\frac{\delta}{\delta\bm{x}_\p}(\bm{S}_\p\cdot\langle\bm{\Omega}_\p\rangle)+e_\p\langle\mathbb{E}\rangle(\bm{x}_\p)\nonumber\\
&\hspace{3em}+\frac{e_\p}{c}\frac{\bm{p}_\p}{\gamma_\p m_\p}\times\langle\mathbb{B}\rangle(\bm{x}_\p)\nonumber\\
&\hspace{5em}+\frac{e_\p}{c}\frac{\delta}{\delta\bm{p}_\p}(\bm{S}_\p\cdot\langle\bm{\Omega}_\p\rangle)\times\langle\mathbb{B}\rangle(\bm{x}_\p)\\
\dot{\mathsf{Z}}_\p&=-\frac{i}{2}(\bm{\sigma}\cdot\langle\bm{\Omega}_\p\rangle)\mathsf{Z}_\p.
\end{align}

The Hamiltonian structure associated with this finite-dimensional system is given by specializing the relevant expressions from Section \ref{finite}. The Hamiltonian is given in Eq.\,(\ref{disc_ham}), which specializes to
\begin{align}
\mathfrak{H}(\bar{\bm{z}},\mathbb{D},\mathbb{A})&=\sum_\p \Big(\gamma_\p m_\p c^2 + \bm{S}_\p\cdot\Upsilon_{\bm{B}\p}\cdot\langle\mathbb{B}\rangle(\bm{x}_\p)\Big)\nonumber\\
&+\frac{1}{8\pi}\int \left|\mathbb{D}+4\pi\sum_\p \Pi_1\Big(\bm{S}_\p\cdot\Upsilon_{\bm{E}\p}\mathcal{W}_{\bm{x}_\p}\Big)\right|^2\,d^3\bm{x}\nonumber\\
&+\frac{1}{8\pi}\int |\mathbb{B}|^2\,d^3\bm{x}.
\end{align}
The Poisson bracket is given in Eq.\,(\ref{disc_poisson}), which has two model-dependent parameters, $\langle J_\p\rangle^{ij}$ and $\langle\mathcal{E}_\p\rangle_i$. These parameters can be computed for the model at hand using Eqs.\,(\ref{disc_symp}) and (\ref{disc_estar}), which give
\begin{align}
(\partial_if)\langle J_\p\rangle^{ij}(\partial_j g)&=\frac{\delta f}{\delta\bm{x}}\cdot\frac{\delta g}{\delta\bm{p}}-\frac{\delta f}{\delta\bm{p}}\cdot\frac{\delta g}{\delta\bm{x}}\nonumber\\
&\hspace{-3em}+\frac{e_\p}{c}\langle \bm{B}\rangle\cdot\frac{\delta f}{\delta\bm{p}}\times\frac{\delta g}{\delta\bm{p}}+\frac{1}{2\hbar}\text{Im}\frac{\delta f}{\delta\mathsf{Z}}^{\dagger}\frac{\delta g}{\delta\mathsf{Z}}\\
\delta{z}^i\langle\mathcal{E}_\p^*\rangle_i[\delta\mathbb{D}]&=4\pi\delta{\bm{x}}\cdot \langle\delta\mathbb{D}\rangle.
\end{align}
The finite-dimensional Poisson bracket is therefore
\begin{gather}
\{\mathfrak{G},\mathfrak{H}\}= \sum_\p \func{\mathfrak{G}}{\bm{x}_\p}\cdot\func{\mathfrak{H}}{\bm{p}_\p}-\func{\mathfrak{G}}{\bm{p}_\p}\cdot\func{\mathfrak{H}}{\bm{x}_\p}\nonumber\\
+\sum_\p\frac{e_\p}{c}\langle\mathbb{B}\rangle(\bm{x}_\p)\cdot\func{\mathfrak{G}}{\bm{p}_\p}\times\func{\mathfrak{H}}{\bm{p}_\p}+\sum_\p\frac{1}{2\hbar}\text{Im}\left(\func{\mathfrak{G}}{\mathsf{Z}_\p}^\dagger\func{\mathfrak{H}}{\mathsf{Z}_\p}\right)\nonumber\\
\hspace{0em}+\sum_\p 4\pi e_\p\left(\func{\mathfrak{G}}{\bm{p}_\p}\cdot \left\langle\func{\mathfrak{H}}{\mathbb{D}}\right\rangle(\bm{x}_\p)-\left\langle\func{\mathfrak{G}}{\mathbb{D}}\right\rangle(\bm{x}_\p)\cdot\func{\mathfrak{H}}{\bm{p}_\p}\right)\nonumber\\
+\int 4\pi c\left(\func{\mathfrak{G}}{\mathbb{D}}\cdot\func{\mathfrak{H}}{\mathbb{A}}-\func{\mathfrak{G}}{\mathbb{A}}\cdot\func{\mathfrak{H}}{\mathbb{D}}\right)\,d^3\bm{x}.
\end{gather}

%%%%%%%%%%%%%%%%%%%%%%%%%%%%%%%%%%%%%%%%%%%%%%%%%%%%%%%%%%%%%%%%%%%
\section{Example: gyrokinetic Vlasov-Maxwell\label{gkvm}}
As a second concrete example of the formalism developed in Sections \ref{continuum} and \ref{finite}, this section will apply the theory to a gyrokinetic Vlasov-Maxwell system. This system couples an ensemble of so-called gyrocenters\cite{Brizard_2007} to small-amplitude electromagnetic fields. It serves as a model for kinetic turbulence in strongly-magnetized plasmas. The gyrokinetic Vlasov-Maxwell system is constructed by applying an asymptotic averaging procedure\cite{Kruskal_1962} to the ordinary Vlasov-Maxwell system. The averaging decouples the rapid motion of charged particles around magnetic field lines from their relatively slow drift along and across them. Due to the asymptotic nature of the averaging procedure, truncation of asymptotic series is a necessary ingredient in the formulation of any computable gyrokinetic model. Thus, there are many different variants of gyrokinetic theory found in the literature.\cite{Tronko_2016} This section will use the manifestly gauge-invariant formulation of gyrokinetics given in Ref.\,\onlinecite{Burby_Tronci_2016}.

The phase space $M$ for an individual gyrocenter is specified in the language of Section \ref{continuum} as
\begin{align}
\bm{m}=v_\parallel,
\end{align}
where $v_\parallel\in\mathbb{R}$ is the gyrocenter parallel velocity. It is also conventional to use the symbol $\bm{X}$ for the spatial location of a gyrocenter, which differs slightly from the notation $\bm{x}$ used in Section \ref{continuum}. The dimension of the internal state space for a gyrocenter is therefore $m=1$, while the dimension of $M$ is $n=3+1=4$. In contrast to Vlasov-Maxwell theory, each gyrocenter species has a discrete index $\sigma$ and a continuous index $\mu$. The discrete index determines a species' charge and mass $e_\sigma,m_\sigma$ while the continuous index gives a species' magnetic moment.  The species ``sum" used in Section \ref{continuum} is therefore the combination of a discrete sum and an integral, $\sum \mathcal{Q}\equiv\sum_\sigma \int_0^\infty \mathcal{Q}\,d\mu.$

The Lagrangian $\ell$ governing the dynamics of an individual gyrocenter is given by Eq.\,(\ref{sin_part}) with
\begin{align}
\theta_i(\bm{z};\bm{B})\dot{z}^i&=\bm{\mathcal{P}}(\bm{X},v_\parallel)\cdot\dot{\bm{X}}\\
K(\bm{z};\bm{E},\bm{B})&=\frac{1}{2}m v_\parallel^2 +\Psi(\bm{X}),
\end{align}
where
\begin{align}
\bm{\mathcal{P}}(\bm{X},v_\parallel)&=m v_\parallel \bm{b}_{\text{eq}}(\bm{X})+\frac{e}{c}\bm{A}_{\text{eq}}(\bm{X})\nonumber\\
&\hspace{1em}+\frac{e}{c}\langle\!\langle\bm{B}(\bm{X}+\lambda\bm{\rho})\times\bm{\rho}\rangle\!\rangle\\
\Psi(\bm{X})&=\mu B_{\text{eq}}(\bm{X})-e\langle\!\langle\bm{E}(\bm{X}+\lambda\bm{\rho})\cdot\bm{\rho}\rangle\!\rangle\nonumber\\
&\hspace{.75em}+2\mu\langle\!\langle\lambda\bm{b}_{\text{eq}}(\bm{X})\cdot \bm{B}(\bm{X}+\lambda\bm{\rho})\rangle\!\rangle
\end{align}
are the gyrocenter kinetic momentum and effective potential. Here $\bm{A}_{\text{eq}}$ is the vector potential of a prescribed, strong, time-independent background magnetic field $\bm{B}_{\text{eq}}=\nabla\times\bm{A}_{\text{eq}}$, while $B_{\text{eq}}=|\bm{B}_{\text{eq}}|$ and $\bm{b}_{\text{eq}}=\bm{B}_{\text{eq}}/B_{\text{eq}}$. The so-called gyroradius vector $\bm{\rho}$ appearing in these expressions depends on $\bm{X}$ and an angular parameter $\theta$ (the gyrophase) according to
\begin{align}
\bm{\rho}&=\rho\,\bm{b}_{\text{eq}}(\bm{X})\times (\cos\theta \bm{e}_1(\bm{X})-\sin\theta \bm{e}_2(\bm{X}))\\
\rho&=\sqrt{\frac{2m c^2\mu  }{e^2B_{\text{eq}}(\bm{X})}},
\end{align} 
where $\bm{e}_{1,2}(\bm{X})$ are mutually orthogonal unit vectors perpendicular to $\bm{B}_{\text{eq}}(\bm{X})$. Finally, the double angle brackets denote averaging over the two parameters $\lambda$ and $\theta$,
\begin{align}
\langle\!\langle\mathcal{Q}\rangle\!\rangle&\equiv \frac{1}{2\pi}\int_0^1\int_0^{2\pi}\mathcal{Q}\,d\theta\,d\lambda.
\end{align}

The definitions of the auxiliary electric and magnetic fields implied by this $\ell$ are, according to Eqs.\,(\ref{cont_D}-\ref{cont_H}),
\begin{align}
\bm{D}(\bm{x})&=\bm{E}(\bm{x})+4\pi\sum \int\langle\!\langle e\bm{\rho}\,\hat{\delta}^3_{\bm{x}}(\bm{X})\rangle\!\rangle\,F\,d^n\bm{z}\label{gvm_D}\\
\bm{H}(\bm{x})&=\bm{B}(\bm{x})+4\pi\sum\int\langle\!\langle[2\mu \lambda\bm{b}_{\text{eq}}(\bm{X})]\,\hat{\delta}^3_{\bm{x}}(\bm{X})\rangle\!\rangle\,F\,d^n\bm{z}\nonumber\\
&\hspace{1em}+4\pi\sum\int\langle\!\langle[ec^{-1}\bm{u}(\bm{z})\times\bm{\rho}]\,\hat{\delta}^3_{\bm{x}}(\bm{X})\rangle\!\rangle\,F\,d^n\bm{z}, 
\end{align}
where $\hat{\delta}^3_{\bm{x}}(\bm{X})=\delta^3(\bm{X}+\lambda\bm{\rho}-\bm{x})$. Likewise, the components of the velocity vector $\bm{V}=(\bm{u},a_\parallel)$ associated with $\ell$ are given by Eq.\,(\ref{cont_hamilton_on_M}), which specializes to
\begin{align}
\bm{u}&=v_\parallel \frac{\bm{B}^*}{B_\parallel^*}+\frac{c\bm{E}^*\times\bm{b}_{\text{eq}}}{B_\parallel^*}\\
a_\parallel&=\frac{e}{m}\frac{\bm{E}^*\cdot\bm{B}^*}{B_\parallel^*},\label{gvm_a}
\end{align}
where
\begin{align}
\bm{B}^*&=\bm{B}+\frac{c}{e}\nabla\times\bm{\mathcal{P}}\\
\bm{E}^*&=\bm{E}-\frac{1}{e}\nabla\Psi-\frac{1}{c}\langle\!\langle\partial_t\bm{B}(\bm{X}+\lambda\bm{\rho})\times\bm{\rho}\rangle\!\rangle\\
B_\parallel^*&=\bm{b}_{\text{eq}}\cdot\bm{B}^*.
\end{align}
The gyrokinetic Vlasov-Maxwell system is then comprised of Eqs.\,(\ref{gvm_D}-\ref{gvm_a}) together with Maxwell's equations in a medium, (\ref{cont_faraday}-\ref{cont_divB}).

In accordance with the general theory developed in Section \ref{continuum}, the gyrokinetic Vlasov-Maxwell system can be cast as an infinite-dimensional Hamiltonian system on $(F,\bm{E},\bm{A})$-space. According to Eq.\,(\ref{cont_ham}), the Hamiltonian functional is given by
\begin{align}
&\mathcal{H}(F,\bm{E},\bm{A})=\sum\int\left(\frac{1}{2}m v_\parallel^2+\mu B_{\text{eq}}(\bm{X})\right)\, F \,d^n\bm{z}\nonumber\\
&+\sum\int2\mu\langle\!\langle\lambda\bm{b}_{\text{eq}}(\bm{X})\cdot\bm{B}(\bm{X}+\lambda\bm{\rho})\rangle\!\rangle\,F\,d^n\bm{z}\nonumber\\
&+\frac{1}{8\pi}\int\left|\bm{D}(\bm{x})-4\pi\sum\int \langle\!\langle e\bm{\rho}\,\hat{\delta}_{\bm{x}}(\bm{X})\rangle\!\rangle\,F\,d^n\bm{z}\right|^2\,d^3\bm{x}\nonumber\\
&+\frac{1}{8\pi}\int |\bm{B}(\bm{x})|^2\,d^3\bm{x}.
\end{align}
The Poisson bracket is given by Eq.\,(\ref{cont_poisson}), which can be written explicitly as
\begin{widetext}
\begin{gather}
\{\mathcal{G},\mathcal{H}\}=\sum\int\frac{1}{m}\frac{\bm{B}^*}{B_\parallel^*}\cdot\bigg(\left\{\func{}{\bm{X}}\func{\mathcal{G}}{F}-e\bm{\mathcal{E}}^*\left[\func{\mathcal{G}}{\bm{D}}\right]\right\}\func{}{v_\parallel}\func{\mathcal{H}}{F}-\func{}{v_\parallel}\func{\mathcal{G}}{F}\left\{\func{}{\bm{X}}\func{\mathcal{H}}{F}-e\bm{\mathcal{E}}^*\left[\func{\mathcal{H}}{\bm{D}}\right]\right\}\bigg)\,F\,d^n\bm{z}\nonumber\\
-\sum\int\frac{c}{e}\frac{\bm{b}_{\text{eq}}}{B_\parallel^*}\cdot\left\{\func{}{\bm{X}}\func{\mathcal{G}}{F}-e\bm{\mathcal{E}}^*\left[\func{\mathcal{G}}{\bm{D}}\right]\right\}\times\left\{\func{}{\bm{X}}\func{\mathcal{H}}{F}-e\bm{\mathcal{E}}^*\left[\func{\mathcal{H}}{\bm{D}}\right]\right\}\,F\,d^n\bm{z}\nonumber\\
+\int4\pi c\left(\func{\mathcal{G}}{\bm{D}}\cdot\func{\mathcal{H}}{\bm{A}}-\func{\mathcal{G}}{\bm{A}}\cdot\func{\mathcal{H}}{\bm{D}}\right)\,d^3\bm{x},
\end{gather}
\end{widetext}
where
\begin{align}
\bm{\mathcal{E}}^*[\delta\bm{D}]=4\pi\delta\bm{D}+4\pi\langle\!\langle(\nabla\times\delta\bm{D})(\bm{X}+\lambda\bm{\rho})\times\bm{\rho}\rangle\!\rangle.
\end{align}
% In order to strike a balance between complexity and applicability, this section will follow the manifestly gauge-invariant formulation of gyrokinetic theory presented in Ref.\,\onlinecite{Burby_Tronci_2016}.

In line with Section \ref{finite}, the gyrokinetic Vlasov-Maxwell system admits a Hamiltonian continuous-time discretization. The single-macroparticle Lagrangian (\ref{lp_nice}) is determined by
\begin{align}
\langle\theta_\p\rangle_i(\bm{z};\bb{B})\dot{z}^i&=\langle\bm{\mathcal{P}}_\p\rangle(\bm{X},v_\parallel)\cdot\dot{\bm{X}}\\
\langle K_\p\rangle(\bm{z};\bb{E},\bb{B})&=\frac{1}{2}m_\p v_\parallel^2+\langle\Psi_\p\rangle(\bm{X})
\end{align}
where
\begin{align}
\langle\bm{\mathcal{P}}_\p\rangle(\bm{X},v_\parallel) &=m_\p v_\parallel \langle\bm{b}_{\text{eq}}\rangle(\bm{X})+\frac{e_\p}{c}\langle\bm{A}_{\text{eq}}\rangle(\bm{X})\nonumber\\
&\hspace{-3em}+\frac{e_\p}{c}\int \langle\!\langle\bb{B}(\bar{\bm{X}}+\lambda\bar{\bm{\rho}}_\p)\times\bar{\bm{\rho}}_\p\rangle\!\rangle\,\mathcal{W}_{\bm{X}}(\bar{\bm{X}})\,d^3\bar{\bm{X}}\\
\langle\Psi_\p\rangle(\bm{X})&=-e_\p\int\langle\!\langle\bb{E}(\bar{\bm{X}}+\lambda\bar{\bm{\rho}}_\p)\cdot\bar{\bm{\rho}}_\p\rangle\!\rangle\mathcal{W}_{\bm{X}}(\bar{\bm{X}})\,d^3\bar{\bm{X}}\nonumber\\
&\hspace{-4em}+2\mu_\p\int\langle\!\langle\lambda\bm{b}_{\text{eq}}(\bar{\bm{X}})\cdot\bb{B}(\bar{\bm{X}}+\lambda\bar{\bm{\rho}}_\p)\rangle\!\rangle\mathcal{W}_{\bm{X}}(\bar{\bm{X}})\,d^3\bar{\bm{X}}\nonumber\\
&+\mu\langle B_{\text{eq}}\rangle(\bm{X})
\end{align}
Here $\bar{\bm{\rho}}_\p$ is $\bm{\rho}_\p$ evaluated at $\bm{X}=\bar{\bm{X}}$. The discrete auxiliary electric and magnetic fields are given by Eqs.\,(\ref{disc_D}-\ref{disc_H}), which simplify to
\begin{align}
\bb{D}(\bm{x})&=\bb{E}(\bm{x})+4\pi\sum_\p e_\p\int\langle\!\langle\bar{\bm{\rho}}_\p\hat{\delta}^3_{\bm{x}}(\bar{\bm{X}})\rangle\!\rangle\,\mathcal{W}_{\bm{X}_\p}(\bar{\bm{X}})\,d^3\bar{\bm{X}}\\
\bb{H}(\bm{x})&=\bb{B}(\bm{x})\nonumber\\
&\hspace{-2em}+4\pi\sum_\p 2\mu_\p\int\langle\!\langle\lambda\bm{b}_{\text{eq}}(\bar{\bm{X}})\hat{\delta}^3_{\bm{x}}(\bar{\bm{X}})\,\rangle\!\rangle\,\mathcal{W}_{\bm{X}_\p}(\bar{\bm{X}})\,d^3\bar{\bm{X}}\nonumber\\
&\hspace{-2em}-4\pi\sum_\p\frac{e_\p}{c}\int\langle\!\langle\bar{\bm{\rho}}\times\dot{\bm{X}}_\p\,\hat{\delta}^3_{\bm{x}}(\bar{\bm{X}})\rangle\!\rangle\,\mathcal{W}_{\bm{X}_\p}(\bar{\bm{X}})\,d^3\bar{\bm{X}}.
\end{align}
Finally, the single-gyrocenter equations of motion are
\begin{align}
\dot{\bm{X}}_\p &=v_\parallel\frac{\langle\bb{B}^*_\p\rangle }{\langle\bb{B}^*_\p\rangle \cdot \langle \bm{b}_{\text{eq}}\rangle}+\frac{c \langle\bb{E}_\p^*\rangle\times\langle\bm{b}_{\text{eq}}\rangle}{\langle\bb{B}_\p^*\rangle\cdot\langle\bm{b}_{\text{eq}}\rangle}\\
\dot{v}_{\parallel\p}&=\frac{e_p}{m_\p}	\frac{\langle\bb{B}_\p^*\rangle\cdot \bb{E}^*_\p}{\langle\bb{B}_\p^*\rangle\cdot \langle\bm{b}_{\text{eq}}\rangle},
\end{align}
where
\begin{align}
\langle\bb{B}^*_\p\rangle&=\langle\bb{B}_{\text{eq}}\rangle+\frac{c}{e_\p}\nabla\times\langle\bm{\mathcal{P}}_\p\rangle\nonumber\\
\langle\bb{E}^*_\p\rangle&=\langle\bb{E}\rangle-\frac{1}{e_\p}\nabla\langle\Psi_\p\rangle\nonumber\\
&-\frac{1}{c}\int\langle\!\langle\partial_t\bb{B}(\bar{\bm{X}}+\lambda\bar{\bm{\rho}})\times\bar{\bm{\rho}}\rangle\!\rangle\mathcal{W}_{\bm{X}}(\bar{\bm{X}})\,d^3\bar{\bm{X}}.
\end{align}
Note that $\langle\bm{b}_{\text{eq}}\rangle$ is \emph{not} usually a unit vector.

The Hamiltonian structure associated with this semi-discrete gyrokinetic system is specified as folllows. The Hamiltonian function (\ref{disc_ham}) is given by
\begin{align}
\mathfrak{H}(\bar{\bm{z}},\bb{D},\bb{A})&=\sum_\p \Big(\frac{1}{2}m_\p v_{\parallel\p}^2+\mu_\p\langle B_{\text{eq}}\rangle(\bm{X}_\p)\Big)\nonumber\\
&\hspace{-3em}\sum_\p 2\mu_\p \int \langle\!\langle\lambda \bm{b}_{\text{eq}}(\bar{\bm{X}})\cdot\bb{B}(\bar{\bm{X}}+\lambda\bar{\bm{\rho}}_\p)\rangle\!\rangle\mathcal{W}_{\bm{X}_\p}(\bar{\bm{X}})\,d^3\bar{\bm{X}}\nonumber\\
&\hspace{-5em}+\frac{1}{8\pi}\int \Big|\bb{D}(\bm{x})-4\pi\sum_\p e_\p\int\langle\!\langle\bar{\bm{\rho}}_\p\,\hat{\delta}^3_{\bm{x}}(\bar{\bm{X}})\rangle\!\rangle\mathcal{W}_{\bm{X}_\p}(\bar{\bm{X}})\,d^3\bar{\bm{X}}\Big|^2\,d^3\bm{x}\nonumber\\
&+\frac{1}{8\pi}\int|\bb{B}(\bm{x})|^2\,d^3\bm{x},
\end{align}
and the Poisson bracket (\ref{disc_poisson}) is given by
\begin{gather}
\hspace{-20em}\{\mathfrak{G},\mathfrak{H}\}=\nonumber\\
\sum_\p\frac{1}{m_\p}\frac{\langle\bb{B}^*_\p\rangle}{\langle\bb{B}^*_\p\rangle\cdot\langle\bm{b}_{\text{eq}}\rangle}\cdot\left(\func{\mathfrak{G}}{\bm{X}_\p}-e_\p \langle\bm{\mathcal{E}}^*_\p\rangle\left[\func{\mathfrak{G}}{\bb{D}}\right](\bm{z}_\p)\right)\func{\mathfrak{H}}{v_{\parallel\p}}\nonumber\\
-\sum_\p\frac{1}{m_\p}\frac{\langle\bb{B}^*_\p\rangle}{\langle\bb{B}^*_\p\rangle\cdot\langle\bm{b}_{\text{eq}}\rangle}\cdot\left(\func{\mathfrak{H}}{\bm{X}_\p}-e_\p \langle\bm{\mathcal{E}}^*_\p\rangle\left[\func{\mathfrak{H}}{\bb{D}}\right](\bm{z}_\p)\right)\func{\mathfrak{G}}{v_{\parallel\p}}\nonumber\\
-\sum_\p\frac{c\langle\bm{b}_{\text{eq}}\rangle}{e_\p\langle\bb{B}_\p^*\rangle\cdot\langle\bm{b}_{\text{eq}}\rangle}\cdot\left(\func{\mathfrak{G}}{\bm{X}_\p}-e_\p \langle\bm{\mathcal{E}}^*_\p\rangle\left[\func{\mathfrak{G}}{\bb{D}}\right](\bm{z}_\p)\right)\nonumber\\
\hspace{8.5em}\times \left(\func{\mathfrak{H}}{\bm{X}_\p}-e_\p \langle\bm{\mathcal{E}}^*_\p\rangle\left[\func{\mathfrak{H}}{\bb{D}}\right](\bm{z}_\p)\right)\nonumber\\
+\int 4\pi c\left(\func{\mathfrak{G}}{\mathbb{D}}\cdot\func{\mathfrak{H}}{\mathbb{A}}-\func{\mathfrak{G}}{\mathbb{A}}\cdot\func{\mathfrak{H}}{\mathbb{D}}\right)\,d^3\bm{x}
\end{gather}
where
\begin{gather}
\langle\bm{\mathcal{E}}^*_\p\rangle[\delta\bb{D}](\bm{z}_\p)=4\pi\delta\bb{D}(\bm{X}_\p)\nonumber\\
+4\pi\int\langle\!\langle\C\delta\bb{D}(\bar{\bm{X}}+\lambda\bar{\bm{\rho}}_\p)\times\bar{\bm{\rho}}_\p\rangle\!\rangle\mathcal{W}_{\bm{X}_\p}(\bar{\bm{X}})\,d^3\bar{\bm{X}}.
\end{gather}

%%%%%%%%%%%%%%%%%%%%%%%%%%%%%%%%%%%%%%%%%%%%%%%%%%%%%%%%%%%%%%%%%%%
%\section{Discrete Hamiltonian structure and simulation verification\label{benefits}}

%\subsection{integral invariants}

%\subsection{partition function}

%%%%%%%%%%%%%%%%%%%%%%%%%%%%%%%%%%%%%%%%%%%%%%%%%%%%%%%%%%%%%%%%%%%
\section{conclusion}
The collisionless kinetic truncation developed in this article is general enough to simultaneously provide Hamiltonian truncations for a variety of important collisionless kinetic plasma models. These models include, but are not limited to, Vlasov-Maxwell, relativistic spin Vlasov-Maxwell, drift-kinetic Vlasov-Maxwell, and gyrokinetic Vlasov-Maxwell. In order to show that all of this generality actually gives something new, Sections \ref{rel_spin} and \ref{gkvm} applied the general theory to derive novel Hamiltonian truncations of the relativistic spin Vlasov-Maxwell system and the gyrokinetic Vlasov-Maxwell system. In the future, it would be interesting to apply the theory to an ensemble of oscillation centers\cite{Dewar_1976} interacting with an electromagnetic field via ponderomotive forces. Such a truncation could be useful in the study of laser-plasma interaction.

The results of this article highlight several open questions, one of which I will mention here. All work done so far on truncating kinetic plasma models while preserving Hamiltonian structure, including the work presented here, uses a PIC discretization to represent the distribution function $F$. Is Hamiltonian truncation compatible with so-called continuum discretizations of $F$? Where PIC discretization is best thought of as a monte carlo sampling of the distribution function, continuum discretization instead uses Galerkin basis functions in phase space to represent $F$, much as in Eulerian fluid simulations. As of now, there are no satisfying answers to this question. This is unfortunate because continuum discretizations are known to suffer much less from discretization noise than PIC discretizations. Enterprising individuals wishing to make an impact in the area of structure-preserving integration may find investigations into this question lucrative.

\section{Acknowledgements}
This research was supported by the U.S. Department of Energy Fusion Energy Sciences Postdoctoral Research Program administered by the Oak Ridge Institute for Science and Education (ORISE) for the DOE. ORISE is managed by Oak Ridge Associated Universities (ORAU) under DOE contract number DE-AC05-06OR23100. All opinions expressed in this paper are the author's and do not necessarily reflect the policies and views of DOE, ORAU, or ORISE
%%%%%%%%%%%%%%%%%%%%%%%%%%%%%%%%%%%%%%%%%%%%%%%%%%%%%%%%%%%%%%%%%%%

\begin{appendix}
\section{Component form of the finite-dimensional kinetic model}
When bases are selected for the discrete spaces of fields, $W_i$, the semi-discrete model presented in Section \ref{finite} can be expressed in a convenient component-based form. In this appendix, I will display this alternative representation of the finite-dmensional model, which is especially convenient for the sake of developing numerical simulations.

Let
\begin{align*}
\{\psi_K\}_{K\in \mathcal{K}}&\subset W_0\hspace{3em}\{\bm{\alpha}_I\}_{I\in\mathcal{I}}\subset W_1\\
\{\bm{\beta}_J\}_{J\in\mathcal{J}}&\subset W_2\hspace{3em}\{\lambda_L\}_{L\in\mathcal{L}}\subset W_3
\end{align*} 
be bases for the spaces $W_i$. As a convenient notational convention,  repeated indices will obey the Einstein summation rule, and the letters $K,I,J,L$ will always be used to index the bases of $W_0,W_1,W_2,W_3$, respectively. Also, for the sake of readability, if $\mathbb{F}$ is an element of $W_i$ for some $i$, the components of $\mathbb{F}$ will be denoted using the corresponding lower-case letter $f$. Thus, the discrete fields $\bb{A},\Phi,\bb{E},\bb{B},\bb{D},\bb{H}$ will be expressed as
\begin{align*}
\bb{A}&=a_I\balpha_I\hspace{3em}\bb{B}=b_J\bbeta_J\hspace{3em}\bb{H}=h_J\bbeta_J\\
\Phi&=\phi_K\psi_K\hspace{2.4em}\bb{E}=e_I\balpha_I\hspace{3.1em}\bb{D}=d_I\balpha_I
\end{align*}

Each $W_i$, by definition, is a subspace of either $L^2(Q)$ or $\bm{L}^2(Q)$. Therefore, each $W_i$ is a real inner product space with inner product given by restricting the appropriate $L^2$ inner product to $W_i$. Let
\begin{align}
\aleph_{K\bar{K}}&=\int \psi_K(\bm{x})\psi_{\bar{K}}(\bm{x})\,d^3\bm{x}\\
\text{I}_{I\bar{I}}&=\int \balpha_I(\bm{x})\cdot\balpha_{\bar{I}}(\bm{x})\,d^3\bm{x}\\
\text{II}_{J\bar{J}}&=\int \bbeta_J(\bm{x})\cdot\bbeta_{\bar{J}}(\bm{x})\,d^3\bm{x}\\
\text{III}_{L\bar{L}}&=\int \lambda_L(\bm{x})\lambda_{\bar{L}}(\bm{x})\,d^3\bm{x}
\end{align}
denote the matrix elements for these inner products. The inverses of these matrices will be denoted $\aleph^{-1}$, $\text{I}^{-1}$, $\text{II}^{-1}$, and $\text{III}^{-1}$. These inner products give rise to the orthogonal projection operators $\Pi_0$, $\Pi_1$, $\Pi_2$, and $\Pi_3$, given by
\begin{align}
\Pi_0(u)&=\aleph^{-1}_{K\bar{K}}\left(\int u(\bm{x})\psi_{\bar{K}}(\bm{x})\,d^3\bm{x}\right)\,\psi_K\\
\Pi_1(\bm{u})&=\text{I}^{-1}_{I\bar{I}}\left(\int\bm{u}(\bm{x})\cdot \balpha_{\bar{I}}(\bm{x})\,d^3\bm{x}\right)\,\balpha_I\\
\Pi_2(\bm{u})&=\text{II}^{-1}_{J\bar{J}}\left(\int \bm{u}(\bm{x})\cdot \bbeta_{\bar{J}}(\bm{x})\,d^3\bm{x}\right)\,\bbeta_J\\
\Pi_3(u)&=\text{III}^{-1}_{L\bar{L}}\left(\int u(\bm{x})\lambda_{\bar{L}}(\bm{x})\,d^3\bm{x}\right)\,\lambda_L.
\end{align}
The inner products also appear in the definition of the functional derivative with respect to elements of the $W_i$. For instance if $\mathfrak{G}(\bb{E},\bb{B})$ is a scalar functional of $\bb{E}\in W_1$ and $\bb{B}\in W_2$, the functional derivatives of $\mathfrak{G}$ are given by
\begin{align}
\frac{\delta\mathfrak{G}}{\delta\bb{E}}&=\text{I}^{-1}_{I\bar{I}}\frac{\partial\mathfrak{G}}{\partial e_{\bar{I}}}\balpha_I\\
\func{\mathfrak{G}}{\bb{B}}&=\text{II}^{-1}_{J\bar{J}}\frac{\partial\mathfrak{G}}{\partial b_{\bar{J}}}\bbeta_J.
\end{align}

According to Eqs.\,(\ref{conform_grad}-\ref{conform_div}), the $W_i$ are well behaved with respect to the standard differential operatos of vector calculus. The discrete $\text{grad}$, $\text{curl}$, and $\text{div}$ operators, which are denoted $\G$, $\C$, and $\D$, may therefore be represented by rectangular matrices $G_{IK}$, $C_{JI}$, and $D_{LJ}$, where
\begin{align}
\G \psi_K&=G_{IK}\balpha_I\\
\C \balpha_I&=C_{JI}\bbeta_J\\
\D \bbeta_J&=D_{LJ}\lambda_L.
\end{align}
Thus, the components of the discrete electric and magnetic field are related to the components of the discrete potentials according to
\begin{align}
e_I&=-\frac{1}{c}\partial_t a_I-G_{IK}\phi_K\\
b_J&=C_{JI}a_I.
\end{align}
The transposed operators $\G^T$, $\C^T$, and $\D^T$ also have matrix representations $(G^T)_{KI}$, $(C^T)_{IJ}$, and $(D^T)_{JL}$, and these are related to the un-transposed matrix elements according to
\begin{align}
(G^T)_{KI}&=\aleph^{-1}_{K\bar{K}} G_{\bar{I}\bar{K}}\text{I}_{\bar{I}I}\\
(C^T)_{IJ}&=\text{I}^{-1}_{I\bar{I}}C_{\bar{J}\bar{I}}\text{II}_{\bar{J}J}\\
(D^T)_{JL}&=\text{II}^{-1}_{J\bar{J}} D_{\bar{L}\bar{J}}\text{III}_{\bar{L}L}.
\end{align}

With all of this notation in place, the finite-dimesional model may now be expressed in component form. The discrete constitutive relations (\ref{disc_D}-\ref{disc_H}) are given by
\begin{align}
d_I&=e_I-4\pi\sum_\p\text{I}^{-1}_{I\bar{I}}\frac{\partial \langle K_\p\rangle}{\partial e_{\bar{I}}}(\bm{z}_\p)\\
h_J&=b_J+4\pi\sum_\p \text{II}^{-1}_{J\bar{J}}\left(\frac{\partial\langle K_\p\rangle}{\partial b_{\bar{J}}}(\bm{z}_\p)-\dot{z}_\p^i\frac{\partial\langle\theta_\p\rangle_i}{\partial b_{\bar{J}}}(\bm{z}_\p)\right).
\end{align}
The discrete Maxwell equations, (\ref{disc_faraday}-\ref{disc_divB}) are
\begin{align}
C_{JI}e_I&=-\frac{1}{c}\partial_t b_J\\
(C^T)_{I J}h_J&=\frac{4\pi}{c}\sum_\p \text{I}^{-1}_{I\bar{I}}\dot{\bm{x}}_\p\cdot\langle\bm{\alpha}_{\bar{I}}\rangle(\bm{x}_\p)+\frac{1}{c}\partial_t d_I\\
(G^T)_{KI}d_I&=-4\pi \sum_\p \aleph^{-1}_{K\bar{K}}\langle \psi_{\bar{K}}\rangle(\bm{x}_\p)\\
D_{LJ}b_J&=0.
\end{align}
Finally, the macroparticle dynamics specified by Eq.\,(\ref{disc_par}) can be written
\begin{align}
\dot{z}^i_\p=\langle J_\p\rangle^{ij}\bigg(\frac{ \partial\langle K_\p\rangle}{\partial z_\p^j}-e_\p \cancel{\delta}^l_j e_I\langle\balpha_I\rangle_l+\partial_t b_J\frac{\partial\langle\theta_\p\rangle_j}{\partial b_J}\bigg).
\end{align}

The Hamiltonian and Poisson bracket for the finite-dimensional model may also be expressed in component form. The Hamiltonian (\ref{disc_ham}) is
\begin{align}
\mathfrak{H}(\bar{\bm{z}},\bb{D},\bb{A})&=\sum_\p\Big(\langle K_\p\rangle(\bm{z}_\p)-\hat{e}_I\frac{\partial\langle K_\p\rangle}{\partial \hat{e}_I}(\bm{z}_\p)\Big)\nonumber\\
&+\frac{1}{8\pi} \hat{e}_I\text{I}_{I\bar{I}}\hat{e}_{\bar{I}}+\frac{1}{8\pi}b_J\text{II}_{J\bar{J}}b_{\bar{J}},
\end{align}
while the Poisson bracket is given by
\begin{align}
\{\mathfrak{G},\mathfrak{H}\}&=4\pi c\, \text{I}^{-1}_{I\bar{I}} \left(\frac{\partial \mathfrak{G}}{\partial d_I}\cdot\frac{\partial \mathfrak{H}}{\partial a_{\bar{I}}}-\frac{\partial\mathfrak{G}}{\partial a_I}\cdot\frac{\partial\mathfrak{H}}{\partial d_{\bar{I}}}\right)\nonumber\\
&+\sum_\p \left(\frac{\partial\mathfrak{G}}{\partial z_\p^i}-e_\p \,\text{I}^{-1}_{I_1\bar{I}_1}\frac{\partial\mathfrak{G}}{\partial d_{\bar{I}_1}}\langle\mathcal{E}_\p^*\rangle_{i I_1}(\bm{z}_\p)\right)\langle J_\p\rangle^{ij}\nonumber\\
&\hspace{1.6em}\times\left(\frac{\partial\mathfrak{H}}{\partial z_\p^j}-e_\p\, \text{I}^{-1}_{I_2\bar{I}_2}\frac{\partial\mathfrak{H}}{\partial d_{\bar{I}_2}}\langle\mathcal{E}_\p^*\rangle_{j I_2}(\bm{z}_\p)\right),
\end{align}
where
\begin{align}
\langle\mathcal{E}_\p^*\rangle_{i I}(\bm{z})=4\pi \cancel{\delta}^j_i\langle\balpha_I\rangle_j(\bm{x})+\frac{4\pi c}{e_\p}C_{JI}\frac{\partial\langle\theta_\p\rangle_i}{\partial b_J}(\bm{z}).
\end{align}

\end{appendix}

%\bibliography{cumulative_bib_file}

%%%%%%%%%%%%%%%%%%%%%%%%%%%%%%%%%%%%%

%% put content of bib file here when ready to submit
\providecommand{\noopsort}[1]{}\providecommand{\singleletter}[1]{#1}%
%

%%%%%%%%%%%%%%%%%%%%%%%%%%%%%%%%%%%%

\end{document}